\documentstyle[preprint,eqsecnum,aps,epsf]{revtex}
\newif\iftightenlines\tightenlinesfalse
\tightenlines\tightenlinestrue
\begin{document}
%
\def\pT{p_T^{\phantom{7}}}
\def\MW{M_W^{\phantom{7}}}
\def\ET{E_T^{\phantom{7}}}
\def\bh{\bar h}
\def\lm{\,{\rm lm}}
\def\lo{\lambda_1}                                              
\def\lt{\lambda_2}
\def\ETC{E_T^c}
\def\pslt{p\llap/_T}
\def\eslt{E\llap/_T}
\def\etmiss{E\llap/_T}
\def\eslt{E\llap/_T}
\def\to{\rightarrow}
\def\Re{{\cal R \mskip-4mu \lower.1ex \hbox{\it e}}\,}
\def\Im{{\cal I \mskip-5mu \lower.1ex \hbox{\it m}}\,}
\def\SU{SU(2)$\times$U(1)$_Y$}
\def\te{\tilde e}
\def\tl{\tilde l}
\def\tb{\tilde b}
\def\tst{\tilde t}
\def\tt{\tilde t}
\def\ttau{\tilde \tau}
\def\tmu{\tilde \mu}
\def\tg{\tilde g}
\def\tga{\tilde \gamma}
\def\tnu{\tilde\nu}
\def\tell{\tilde\ell}
\def\tq{\tilde q}
\def\tw{\widetilde W}
\def\tz{\widetilde Z}
\def\cmsec{{\rm cm^{-2}s^{-1}}}
\def\fb{{\rm fb}}
\def\sgn{\mathop{\rm sgn}}
\def\mhf{m_{\frac{1}{2}}}
\def\jet{{\rm jet}}
\def\jets{{\rm jets}}

\hyphenation{mssm}
\def\ds{\displaystyle}
\def\ts{${\strut\atop\strut}$}
%
%
\preprint{\vbox{\baselineskip=14pt%
   \rightline{FSU-HEP-960415}\break 
   \rightline{UCD-96-11}\break
   \rightline{UH-511-847-96}
}}
\title{SUPERSYMMETRY REACH OF TEVATRON UPGRADES:\\
A COMPARATIVE STUDY}
\author{Howard Baer$^1$, Chih-hao Chen$^2$, 
Frank Paige$^3$ and Xerxes Tata$^{4}$}
\address{
$^1$Department of Physics,
Florida State University,
Tallahassee, FL 32306, USA
}
\address{
$^2$Davis Institute of High Energy Physics,
University of California,
Davis, CA 95616, USA
}
\address{
$^3$Brookhaven National Laboratory, 
Upton, NY 11973, USA
}
\address{
$^4$Department of Physics and Astronomy,
University of Hawaii,
Honolulu, HI 96822, USA
}
\date{\today}
\maketitle
\begin{abstract}

We use ISAJET to perform a detailed comparison of the supersymmetry
reach of the current Tevatron (100~pb$^{-1}$)
with that of the Main Injector (2~fb$^{-1}$) and
the proposed TeV33 upgrade designed to yield an integrated
luminosity of 25~fb$^{-1}$. Our analysis is performed within the
framework of the minimal supergravity model with gauge coupling
unification and radiative electroweak symmetry breaking.
For each of these three luminosity options, we delineate
the regions of parameter space where jets plus missing
energy plus 0, 1, 2 (opposite sign and same-sign dileptons), and 3
isolated lepton signals from the cascade decays of sparticles
should be visible above standard model
backgrounds. We compare these with the parameter regions where signals in the 
clean isolated dilepton and trilepton channels
(from chargino/neutralino and slepton production) should be observable.
\end{abstract}

\medskip



\section{Introduction}

The CDF and D0 experiments have each
accumulated an integrated luminosity of
about 100~pb$^{-1}$ in Run I of the Fermilab Tevatron.
An analysis of these data, which include elementary particle
collisions at the highest energies accessible today, has already led
to the discovery of the top quark\cite{TOP}, and could reveal
deviations from the expectations of the Standard Model (SM) of particle
physics. These could take the form of new degrees of freedom (supersymmetry,
technicolor, new gauge bosons, \ldots) or new effective interactions
(quark compositeness, \ldots).
At the very least, if no such deviation is observed, these data would
serve to put phenomenological bounds on various extensions of the SM.
Around 1999, the Main Injector (MI) is expected to begin operation: this should
result in an order of magnitude increase in the size of the
data sample. Tevatron
experiments should then be sensitive to new physics processes with
cross sections that are ten (three) times smaller than those 
accessible
from an analysis of 
Run I data in rate-limited
(background-limited) channels.

Rather general arguments based on
the instability of the SM's electroweak-breaking sector 
to high mass scales suggest
that it should break down at an energy scale $\Lambda \alt 1$~TeV,
which is the {\it raison d'\^etre} for supercolliders such as the
Large Hadron Collider (LHC) at CERN, or an $e^+e^-$ collider operating
at a center of mass energy of $\sim$1~TeV. If $\Lambda$ is well below its 
upper bound,
deviations from the SM may also manifest themselves at the Tevatron, 
if sufficient integrated luminosity
can be accumulated. Recently, the demise of the
Superconducting Supercollider (SSC) has led several authors\cite{TEV2K}
to propose that a luminosity upgrade beyond the Main Injector,
designed to provide a data sample of 10~--~25~fb$^{-1}$, could have a
significantly improved reach for new physics well before the LHC
commences operation. We will refer to this proposed upgrade as TeV33.
We make no attempt here to assess the credibility of the TeV33 goals,
either for the accelerator or for the existing or upgraded detectors.

The purpose of this paper is to make a quantitative comparison between
the capabilities of the current Tevatron, the MI upgrade, and the proposed
TeV33 for the discovery of supersymmetry\cite{REV,DPF}.
We focus on supersymmetry 
for several reasons. Supersymmetry (SUSY) provides the only known
weakly coupled (and hence perturbatively calculable) framework
that can naturally stabilize the Higgs sector of the SM. Since
supersymmetry is a decoupling theory, SUSY models reduce to the SM
if sparticles are heavy: thus, SUSY models are at least as consistent
as the SM when confronted with the precision data from LEP. SUSY models
include a natural candidate for dark matter, and they can be consistently
and simply embedded in a grand unified framework. A completely different
reason for focusing on supersymmetry for the purpose of comparing
the capabilities of different experimental facilities stems from the 
fact that SUSY models contain several new particles with a variety
of quantum numbers: colored scalars and fermions, as well as corresponding
colorless particles. A supersymmetry skeptic could simply view 
our studies as providing a ``theoretical
laboratory'' to compare the capabilities of various projected experimental
facilities.

Several studies of the SUSY reach
of TeV33 already exist\cite{LOPEZ,BCKT,KANE}, and results
have been summarized in Ref.\cite{DPF,TEV2K}. 
These studies do not all focus on the same SUSY reactions, 
differ significantly 
in the details of the computation of backgrounds (crucial in
determining the reach), and present the final results in forms not
amenable to direct comparison\cite{FN1}. In view of the importance 
of this issue, we felt that a systematic study in which all signals
are simultaneously studied using a common simulation would be useful
in making an assessment of the increased capability of TeV33 over the
already approved MI upgrade. Moreover, we present for the first
time a comparison
of the reach in various multilepton channels for the three 
luminosity options at the Tevatron.

For definiteness, we work within the minimal supergravity framework,
with assumptions 
(including grand unification and radiative electroweak symmetry breaking)
detailed in our earlier studies\cite{BCMPT,BCKT} of 
SUSY signals at the Tevatron. The masses and couplings of all sparticles
are then determined by just SUSY four parameters,
\begin{itemize}
\item $m_0$, the common scalar mass at the unification scale,
\item $\mhf$, the common gaugino mass at the unification scale,
\item $A_0$, the common soft SUSY breaking trilinear scalar coupling, and
\item $\tan\beta$, the ratio of the vacuum expectation values of the two Higgs
fields,
\end{itemize}
together with the sign of the Higgsino mass term $\mu$. Our results
also depend on $m_t$; 
we shall take $m_t=170~$GeV. The physical
masses and couplings relevant to phenomenology are then 
obtained\cite{BCMPT,BCKT} using the
renormalization group equations. We emphasize that we use this framework
for expediency. Indeed the assumptions underlying this framework may
ultimately prove to be incorrect. This is largely irrelevant for
our purpose, which is only to compare the SUSY reach of the different Tevatron
upgrades. 

To orient the reader with the masses of various sparticles within this 
framework, we show, in Fig.~1, 
contours of the gluino (solid lines), squark (dashed lines)
and lighter chargino (dotted lines) masses
in the $m_0$--$\mhf$
plane. We fix $A_0=0$ and illustrate the masses for
{\it a})~$\tan\beta=2, \mu < 0$, 
{\it b})~$\tan\beta=2, \mu > 0$,
{\it c})~$\tan\beta=10, \mu < 0$, 
and {\it d})~$\tan\beta=10, \mu > 0$. The gluino mass contours are not
exactly horizontal because 
of the difference between the pole and running gluino masses\cite{POLE}.
The bricked regions are excluded by theoretical constraints detailed
in Ref.\cite{BCMPT} while the hatched region is excluded from the
non-observation of any SUSY signal at the Tevatron\cite{TEVLIM}, or
LEP\cite{LEPLIM}, and includes the recent mass limit\cite{RECENT} 
$m_{\tw_1}>65$~GeV from LEP1.5. 

The cascade decays of gluinos and squarks
can lead to a variety of multijet plus multilepton event topologies.
Contributions from associated production processes, while included, are
known to be small.
In addition, electroweak production of charginos and neutralinos can
lead to hadronically quiet multilepton signals via which to search
for supersymmetry. In this paper, we use ISAJET 7.16 \cite{ISAJET}
to map out the supersymmetry reach
in various channels at a 2 TeV $p\bar p$ collider, assuming
an integrated luminosity of
\begin{itemize}
\item 100~pb$^{-1}$, roughly corresponding to what might be achieved from
an analysis of the data from the current run;
\item 2~fb$^{-1}$, corresponding to the approved MI upgrade; and
\item 25~fb$^{-1}$, corresponding to what might be possible at the proposed
TeV33 upgrade.
\end{itemize}
We assume identical detectors for each of these options. 
We neglect event pile-up effects, which have been shown to be small at least
for the clean trilepton channel from $\tw_1\tz_2$ production\cite{TEV2K}.
We also neglect any experimental difficulties associated with
operating at the high luminosities implied by TeV33.
Our purpose here is mainly to evaluate which regions of SUGRA parameter
space can be explored via various discovery channels, and how these
change depending on different Tevatron luminosity options.

The remainder of this paper is organized as follows. In the next section,
we briefly describe the event simulation
that we use for the computation of various signals
and backgrounds. In Sec.~III, we focus on various multijet plus multilepton
signals from the production and cascade decays of
gluinos and squarks as well as from chargino and neutralino production,
and map out the reach for supersymmetry in several of
these channels for the three luminosity options introduced
above. We present our results in the $m_0-\mhf$ plane
for cases ({\it a--d}) introduced above. 
Variation with $A_0$ is briefly addressed.
Sec.~IV is devoted to the study of the corresponding
reach via electroweak production of charginos and neutralinos.
We compare the total SUSY reach for the three Tevatron options and briefly
discuss how these compare to the reach for other facilities such as the
LHC or an $e^+e^-$ linear collider operating at 500-1000~GeV in Sec. V.  

\section{Event Simulation}

The implementation of the SUGRA framework into
ISAJET has been described elsewhere\cite{BCMPT} and will
not be repeated here. We generate all the lowest order $2 \to 2$
SUSY subprocesses in our simulation of $n$ lepton plus $m$ jet
signals with $m \geq 2$
(except for 
unimportant $s$-channel Higgs boson mediated subprocesses). However,
for the simulation of the {\it clean} multilepton signals,
we have generated only slepton and chargino/neutralino events, 
since gluino and squark decays will very seldom yield final states 
without central jet activity.

To model the experimental conditions at the Tevatron,
we use the toy calorimeter
simulation package ISAPLT. We simulate calorimetry covering
$-4<\eta <4$ with cell size $\Delta\eta\times\Delta\phi =0.1\times
0.0875$. We take the hadronic (electromagnetic) energy resolution to be $70\%
/\sqrt{E}$ ($15\% /\sqrt{E}$).
Jets are defined as hadronic clusters with $E_T > 15$~GeV within
a cone with $\Delta R=\sqrt{\Delta\eta^2
+\Delta\phi^2} =0.7$. We require that $|\eta_j| \leq 3$.
Muons and electrons are classified as isolated if they have
$p_T>7$~GeV and
$|\eta (\ell )|<2.5$ and if the visible activity within a cone of $R
=0.3$ about the lepton direction is less than $E_T({\rm cone})=5$ GeV.
In our analysis, we neglect multiple scattering effects and non-physics
backgrounds from photon or jet misidentification, and we make no attempt
4to explicitly simulate any particular detector.

\section{Reach in Multilepton plus Multijet Channels}

The multijet plus $\etmiss$ signal from gluino and squark production
has been regarded as the most promising
signature for supersymmetry at hadron colliders, provided that they
are kinematically accessible. From a non-observation
of any excess of $\eslt$ events above SM expectations, the CDF and D0
experiments\cite{TEVLIM} have already obtained the lower limits
in the vicinity of 100-200~GeV on the masses of gluinos and squarks.
These bounds have been derived within the framework of the
minimal supersymmetric model, assuming ten degenerate squark flavors.
It has also been shown\cite{BTW,BGH,DPROY,RPV}
that depending on $m_{\tq}$ and $m_{\tg}$,
other signatures with
multilepton plus jets in the final state may also be observable with
a data sample of $\cal{O}$(100)~pb$^{-1}$. 

If gluinos and squarks are relatively heavy, electroweak production of
charginos ($\tw_1$) and neutralinos ($\tz_2$) (which are expected
to have masses $\sim \frac{1}{3} m_{\tg}$ in models with a common
gaugino mass at the unification scale) may offer the best hope for
SUSY detection at the Tevatron\cite{BCKT}. With the integrated luminosity
that has been accumulated in Run I of the Tevatron, these signals
should be on the verge of observability\cite{AN,BT,BKT,D0WINO}.
For an integrated luminosity
in excess of $\cal$O(1~fb$^{-1}$) that should be available once the 
MI commences operation, the reach via the clean trilepton signal
from the process $pp \to \tw_1\tz_2 \to \ell\nu\tz_1+\ell'\bar{\ell'}\tz_1$
may exceed that from gluino and squark reactions\cite{LOPEZ,BCKT,KANE}
for a wide range of parameters.
We will defer the discussion of these reactions to the next section, 
and focus our attention on the multijet plus multilepton signatures
for now.

We require,
\begin{itemize}
\item jet multiplicity $n_\jet \geq 2$,

\item transverse sphericity $S_T > 0.2$,

\item $\etmiss > 40$~GeV.
\end{itemize}
As in our analysis of signals at the LHC\cite{LHC}, we require
an analysis cut,
\begin{itemize}
\item $E_T(j_1), \  E_T(j_2) \ > \ E_T^c$ and $\eslt > E_T^c$,
\end{itemize}
where the parameter $\ETC$ is appropriately adjusted as described below.
We further classify the events by their isolated lepton content as follows:
\begin{itemize}
\item  (A) $\etmiss$ events, with no isolated leptons. For this sample, we
require that the missing energy not point along a jet,
$\Delta\phi(\vec{\etmiss},\vec{E_{Tj}})>30^o$;
\item (B)  $1\ell$ events with exactly one isolated lepton with
$E_T(\ell)> 10$~GeV.
To reduce the background from $W$ production, we also
require $M_T(\ell,\etmiss)> 100$~GeV;
\item (C)  Opposite sign ($OS$) dilepton events with exactly two unlike
sign isolated leptons, where we require $E_T(\ell_1)> 10$~GeV;
\item (D)  Same sign ($SS$) dilepton events with exactly two same sign
isolated leptons, again with $E_T(\ell_1)> 10$~GeV;
\item (E)  $3\ell$ events, with exactly three isolated leptons with $E_T(\ell_1)>
10$~GeV. We veto events with $|M(\ell^+\ell^-)-M_Z|<8$~GeV.
\end{itemize}

SUSY events are frequently rich in central $b$ jets which
can come from
direct decays of gluinos and $b$-squarks;
from the decay $\tz_2\to b\bar{b}\tz_1$,
which can have an enhanced branching fraction\cite{BENH}; or
from the production of Higgs bosons in SUSY events.
We have therefore studied the prospects for the detectability
of tagged $b$ signatures in the sample of $\etmiss$ and $1\ell$ SUSY
samples. In our analysis, we take the efficiency of tagging
a $b$-jet with $E_T>15$~GeV and $|\eta_b|<2$ is 50\%,
and assume that the probability of mistagging other jets as
a $b$-jet is 2\%. We find that
the signals for $b$-tagged $1\ell$ events occur at unobservably
small rates over most of the parameter space, so that for the
most part, we will confine our attention to just
\begin{itemize}
\item (F) $B+\jets+\etmiss$ events, where we veto any leptons,
and require at least one tagged $B$-hadron.
\end{itemize}
We have used ISAJET to compute these signals and the corresponding
SM backgrounds which mainly come from,
\begin{itemize}
\item $W$ or $Z$ production, in association with jets (additional
leptons could arise from radiation),
\item $t\bar{t}$ production,
\item $WW$, $WZ$ and $ZZ$ pair production, where jets come from QCD
radiation, and
\item QCD jet production, with $\etmiss$ due to mismeasurement of 
the jets.
\end{itemize}

Some care must be exercised in the generation of the backgrounds. It is
highly more likely that events with a hard initial scattering will
pass the hard cuts that we have imposed to separate the SUSY signal
from the background (especially for large values of $\ETC$). Since the
cross section falls rapidly with the hard scattering $p_T$, it is
necessary to generate the backgrounds in various $p_T$ bins to ensure
that the relevant portions of the phase space are properly sampled.
If this is not done, most of the events generated by the Monte Carlo
procedure will be too soft to pass the hard cuts, 
leading to an underestimate of the
backgrounds. To avoid this we follow the procedure detailed in
Ref.\cite{LHC}, and generate background events in hard scattering
$p_T$ in geometrically increasing bins between 25-400~GeV.

The $\ETC$ dependence of the various backgrounds is shown
in Fig.~2 for ({\it a})~$\etmiss$ events, ({\it b})~$1\ell$~events,
({\it c})~$OS$ dilepton events, ({\it d})~$SS$ dilepton events,
({\it e})~$3\ell$ events, and ({\it f})~$b+\jets+\etmiss$ events.
We use CTEQ2L structure functions\cite{CTEQ} throughout our analysis.
The long-dashed line shows the background from $t\bar{t}$ production,
while the QCD background is shown by the long-dashed-dotted line.
The long-dashed double-dotted (triple-dotted) line shows the backgrounds
from $W+\jets$ ($Z+\jets$). Backgrounds from vector boson pair production
are shown by the short-dashed (dotted) lines and are negligible, since
the cross section for producing vector boson pairs together with at
least two jets is rather small. The sum of all the backgrounds is
shown as the solid line. We see from Fig.~2 that for all but the
$\etmiss$ signal (and, to a lesser extent, the $\jets +B+\eslt$ signal), 
$t\bar{t}$ and $W$+jets backgrounds dominate over
most of the range of $\ETC$. For the $\etmiss$ signal, the background
from $Z \to \nu\bar{\nu}$ plus jet events is also significant \cite{FNKANE} 
and, in fact,
dominates for very large values of $\ETC$. Backgrounds from QCD and 
vector boson pair production are negligible in this case.

We have shown in our previous analyses\cite{LHC} of SUSY
signals at the LHC that the reach can be optimized by adjusting
the value of $\ETC$ in the analyses. Events from relatively
light gluinos and squarks tend to be softer but have larger
cross sections than SUSY events when gluinos and squarks are
quite heavy. While a modest value of $\ETC$ is probably
optimal for the former case, the signal to background ratio as well as
the statistical significance of the signal can
be significantly improved by using larger values of $\ETC$ when attempting
to extract a signal for very heavy sparticles. This should also
be true of experiments at the Tevatron. As the size of the data sample
increases, generally speaking, we may expect that we would obtain
the maximum reach by increasing the value of $\ETC$. The analysis
should, however, {\it not} be done for a single choice of this cut
parameter that is optimized for the maximal reach because this would
result in a very low efficiency for signal detection if sparticles happened
to be relatively light\cite{LHC}.
The optimal choice of $\ETC$ is also channel-dependent and
somewhat sensitive to where we are in the parameter space. 

To underscore this, we have shown
in Fig.~3 are total signal cross sections as a function
of $\ETC$ for the same event
topologies ({\it a})-({\it f}) as in Fig.~2, but for several cases
of input SUGRA parameters. In our illustration, we have fixed
$\mhf=120$~GeV so that $m_{\tg}$ is roughly at 350-380~GeV. 
In the first three cases, we choose $\tan\beta=2,\ A_0=0, \ \mu<0$,
and take 
\begin{itemize}
\item (1) $m_0=100$~GeV (solid)
\item (2) $m_0=200$~GeV (dashed)
\item (3) $m_0=800$~GeV (dotted)
\end{itemize}
Increasing $m_0$ mainly increases the mean squark mass:
in case 1, the decays $\tg\to q\tq$ are kinematically accessible, and
of course, the production of squarks is substantial; in case 2, squarks
of the first two generation are just heavier than the gluino, so that the
dominant gluino decay is via $\tg\to b\tb$; in case 3, all squarks are
very heavy, and the gluino decays via three body modes. We have checked
that flipping the sign of $\mu$ does not qualitatively change the results.
In the remaining two cases 4 and 5,
we fix $m_0=200$~GeV with $\mu>0$, and choose 
\begin{itemize}
\item (4) $A_0=-400$~GeV (dot-dashed)
\item (5) $A_0=-430$~GeV (dot-dot-dashed)
\end{itemize}
These have been
chosen so that the decay $\tg \to t\tt_1$ is kinematically accessible: In case
4, $\tt_1$ decays via $\tt_1\to b\tw_1$, while in case 5 this channel
is inaccessible and $\tt_1$ decays via the loop mode
$\tt_1 \to c\tz_1$ \cite{HK,BDGGT}. The size of the tagged $b$-jet cross
section is worth noting. The wiggles in the curve reflect the statistical
fluctuations in the simulations.
The total SM background is shown by the crosses.
We see that if squarks are heavy (case 3, dotted) the background substantially
exceeds the signal in all the channels and for the entire
$\ETC$ range where the  signal cross section remains observable,
and it appears unlikely
that gluinos (whose mass is 380~GeV) will be detectable even at TeV33.
We also see that for some signals ({\it e.g.} the OS channel), the
choice of $\ETC$ significantly changes the signal to background ratio
depending on the decay pattern of the gluino (compare cases 1 and 2 with 4 
and 5). Clearly some optimization should be possible if the very
large data sample as envisioned at TeV33 become available. 

In order to assess the observability of the various SUSY signals
at the Tevatron for integrated luminosities of 0.1~fb$^{-1}$, 2~fb$^{-1}$
and 25~fb$^{-1}$, we have generated SUSY events for a grid of points
in the $m_0-\mhf$ plane, for $\tan\beta=2$ and $\tan\beta=10$ for
both signs of $\mu$ with $A_0$ being fixed at zero. Our scan should
be taken as spanning a representative portion of the parameter space
for small and modest values of $\tan\beta$\cite{FN2}. Since $A_0$ mainly
affects the phenomenology of the third generation, we expect the signals
to be relatively insensitive to variation of $A_0$ (except for signals
involving third generation fermions, such as the cross
section for $\etmiss$ events with tagged $b$ jets. We will return to this
point later. We have then run the generated SUSY events through our
toy detector simulation and classified them into the various topologies
A--F introduced above. Finally, for each SUGRA point, we have checked
whether the SUSY signals are observable above SM backgrounds for each
of the three values of integrated luminosities. Here, we consider
a signal to be observable if, for the given integrated luminosity,
we have ({\it i}) at least 5 signal events, ({\it ii}) the statistical
significance of the signal exceeds ``5$\sigma$'', {\it i.e.}
$N_{\rm signal} > 5\sqrt{N_{\rm back}}$, and 
({\it iii})~$N_{\rm signal} > 0.2 N_{\rm back}$.
The third requirement, while somewhat arbitrary, is to ensure that we do
not classify the $\etmiss$ signal with $\ETC=25$~GeV (see Fig.~3{\it a})
with a cross section of 300~fb to be observable
above the background of 3$\times 10^4$~fb.
It seems evident to us that because there is no characteristic kinematic
distribution whose shape the signal would grossly distort, this
``1\% excess'', while ``observable'' according to criteria ({\it i})
and ({\it ii}), would be impossible to detect. We check the
observability of SUSY events for $\ETC=15, 40, 60, 80, 100,120$ and 140~GeV
and consider the signal to be observable if it is so for
{\it any one} of the values of $\ETC$.

The results of our computation are illustrated in Fig.~4--Fig.~9, for
each of the event topologies A--F, respectively for the same four choices
of other SUGRA parameters as in Fig.~1. The theoretically and
experimentally excluded regions, denoted by bricks and hatches, are
identical to Fig.~1. 
The regions of the $m_0-\mhf$ plane where a
signal is observable are shown for an integrated luminosity of
100~pb$^{-1}$ (black squares) corresponding to the data sample of
Run I, of 2~fb$^{-1}$ (gray squares) corresponding to a data
sample that might be available at the MI, and finally, of 25~fb$^{-1}$ 
(white squares)
corresponding to a data sample that might become available after a few
years of operation of the proposed TeV33 upgrade.

Several comments are worth noting.
\begin{itemize}
\item By comparing the distribution of the solid squares in Fig.~4--Fig.~9,
we conclude that the $\etmiss$ channel
should provide the maximal reach when the data of Run I is analyzed.
Values of $\mhf$ up to about 100~GeV should be accessible if $m_0 \leq
200$~GeV. For positive values of $\mu$ this region would already have
been accessible via the chargino search at LEP1.5 (from Fig.~1 it is
clear that $\tw_1$ tends to be heavier for $\mu < 0$, with all other
parameters the same) which is why we have no solid squares in Fig.~4{\it b}
and Fig.~4{\it d}. We stress, however, that this depends crucially on the
assumed gaugino mass unification, and a direct for the gluino
signal (even for parameters in the hatched regions) is extremely
important. Also, note from Fig.~4{\it a} that Tevatron experiments
may probe regions of parameter space not accessible at LEP2.

\item It is interesting to see that even with an integrated
luminosity of 100~pb$^{-1}$, we see that there are regions of
the parameter space where there should be confirmatory signal also in other
channels. Except for isolated corners in the small $m_0$ region (where
the leptonic branching fraction of charginos and neutralinos may be enhanced
because their two-body leptonic decays are accessible), the leptonic
or tagged-$b$ channels are unlikely to be discovery channels for supersymmetry
at least for Run I of the Tevatron.

\item With a data sample of 2~fb$^{-1}$ Tevatron experiments should be
able to probe $\mhf$ values up to 150~GeV (corresponding to
$m_{\tg} \sim 400$~GeV) in the $\etmiss$ channel
if $m_0 \alt 200$~GeV. This is considerably beyond the reach of LEP2.
At the high values of $\mhf$ that should be explorable at the MI, it is worth
noting that electroweak production of charginos and neutralinos
is a significant contributor to even the $\etmiss$ channel: the
production of gluinos and squarks is kinematically suppressed, so
that the signals from the electroweak production of charginos and neutralinos
which have masses $\sim (\frac{1}{6}-\frac{1}{3})m_{\tg}$ become independent.
The same is true in the leptonic channels --- in the $3\ell$ channel, the 
jets then mainly come from QCD radiation, which is included in the shower
approximation in ISAJET.
For larger values of $m_0$,
squarks become inaccessible, and the range of $\mhf$ that might be probed
becomes smaller. Not only is the reach at the MI considerably larger, we
see that there should be confirmatory signals in the various leptonic
channels (particularly if $m_0$ is not very large) and even in the tagged
$b$-channel. Once again, we see that the maximum reach is obtained 
in the $\etmiss$ channel, and that the trilepton channel enhances the
reach only for isolated values of SUGRA parameters.  

\item We see from Figs.~4 that TeV33 should be able to probe $\mhf$ values
about 25~GeV beyond what may be explored at the MI in the $\eslt$ channel;
this corresponds to an increase in the gluino mass reach of about 65~GeV.
The increase in the reach may be somewhat larger in the leptonic channels,
as can be seen from Figs.~5--8.
However, the maximal reach is still obtained via the $\eslt$ channel. We
thus conclude that if Tevatron experiments are able to accumulate
about 25~fb$^{-1}$ of data, they would be able to probe gluino masses
about 20-25\% beyond what might be probed 
at the MI ($m_{\tg}\simeq 300-500$ GeV). It is, however,
important to note that with such a large data sample, one would be almost
guaranteed to see a signal in several channels over much of the parameter
range where there is a signal in the $\eslt$ channel.

\item Turning to the prospects for identifying tagged $b$-jets in the
sample of $\eslt$ events, we see from Fig.~9 that it is unlikely from
an analysis of the data from Run I. $b$-tagged $\eslt$
events should be identifiable
over a significant range of parameters at both the MI or TeV33. Measurement
of the tagged $b$ cross section, as well as the multiplicity of $b$-jets
in SUSY events, could serve to yield information about the $A$-parameter
or the $b$-quark Yukawa coupling, if a sufficient number of SUSY events
can be accumulated. We warn the reader that the observability of this signal
is rather sensitive to the assumptions about $b$-tagging and mistagging.
We have checked that the signal to background ratio becomes {\it smaller}
if the mistagging probability is taken to be zero instead of 2\%; {\it i.e.}
the signal contains a larger fraction of mistagged events than the background.
This is mainly due to the fact that the main SM background to the $b$-tagged
sample comes from top quark production, and the background rate from
misidentification of QCD jets in $W$ and $Z$ events adds little
to the main physics background. In addition,
the larger jet multiplicity, and the resulting
greater probability of mistagging jets in the signal sample should lead to
a larger ``fake'' background for the signal events relative to SM events.  

\end{itemize}

Up to now, we have fixed the soft breaking parameter $A_0=0$. In order to
investigate the sensitivity of the cross sections to variation of
$A_0$ we show the Tevatron cross sections after cuts in the six channels
A--F in Fig.~10. We fix $m_0=100$~GeV and $\mhf=120$~GeV, which yields
$m_{\tq} \simeq m_{\tg}-30$~GeV $\simeq 315$~GeV for $A_0=0$. The
masses of the gluino and the first two generations of squarks are only
very weakly dependent on $A_0$. Thus, for the cases studied in Fig.~10,
$\tq\tq,\tq\tg$ and $\tg\tg$ processes all contribute to the signals,
and gluinos decay to squarks. We examine values of $A_0$ for which the
squared stop masses are positive. The following features are worth noting:
\begin{enumerate}
\item In case ({\it a}) the various topological cross sections are
quite insensitive to $A_0$. As we vary $A_0$ over the complete range,
sparticle masses do not change very much, and no thresholds for new
decays are encountered. The gluinos and squarks each decay via two
body modes, with branching fractions relatively insensitive to $A_0$.

\item In case ({\it b}), as we consider decreasing values of $A_0$, starting
with large positive values, the cross sections are relatively constant
until $A_0$ becomes negative enough so that $\tt_1$,
the lighter of the $t$-squarks, becomes accessible via gluino decays.
The scalar top decays via $\tt_1 \to b\tw_1$, so that the number of $b$-jets
in gluino events is significantly increased. Direct production of
$\tt_1\bar{\tt_1}$ pairs also contributes to the increase in the tagged
$b$-jet cross section and should be independently
detectable\cite{SENDER} if $m_{\tt_1}$ is not too large. 
The increased cross sections in the OS and SS lepton
channels also have their origin in $\tg \to t\tt_1$ decays
followed by the semileptonic decays of the top family; $\tt_1\bar{\tt_1}$
production contributes to the increase in the OS as well as the $\eslt$
cross sections.

\item Cases ({\it c}) and ({\it d}) show roughly similar features as case
({\it b}) --- as $A_0$ decreases from large positive values, the cross
sections again show an increase because $\tt_1$ becomes relatively light.
However, a new feature is now seen for large negative values of $A_0$ in case
({\it c}) ---
the $\eslt$ cross section shows a sharp increase, while all other
cross sections drop sharply. This occurs when $\tt_1$ becomes lighter
than the chargino, so that the loop decay mode $\tt_1 \to c\tz_1$ dominates
$\tt_1$ decays, leading to a drop in the $b$-jet multiplicity from
gluino and stop decays. The leptonic cross sections decrease because
the chargino now decays via $\tw_1 \to b\tt_1$ with a branching fraction
of essentially 100\%; because $|m_{\tw_1}-m_{\tt_1}|$ is small, the $b$-jets
are presumably too soft to satisfy the tagging requirements. If SUSY parameters
happen to be in this range, the promising clean trilepton signal from
$\tw_1\tz_2$ production will be absent, and $\eslt$ events will be the
main signature for chargino pair production. Finally, we remark that in
case ({\it d}), $|m_{\tw_1}-m_{\tt_1}|$ is somewhat larger (at extreme
negative values of $A_0$) and the $b$-jet appears to be sufficiently hard.
As a result, while the leptonic cross sections exhibit a sharp decrease,
the $\eslt$ and tagged $b$ topologies do not. 

\end{enumerate}

5\section{Reach via Clean Multilepton Channels}

If gluinos and squarks are beyond the reach of the Tevatron,
it is possible that SUSY might manifest itself through
electroweak processes, via events with two or
more hard isolated leptons and $\eslt$ but with essentially no jet
activity. We will refer to these as clean multilepton channels.
The spectacular trilepton event signature from the reaction
$p\bar{p} \to \tw_1\tz_2 \to \ell\nu\tz_1+\ell'\bar{\ell'}\tz_1$
has recently been the focus of considerable attention\cite{LOPEZ,BCKT,KANE}
but observable signals may also be present in dilepton channels\cite{BCKT}. 
Sources
of clean dilepton events include chargino and slepton pair production.
Since we have studied these channels in detail elsewhere, here
we will mostly focus on the comparison of the capabilities of the
three luminosity options, and refer the reader to our earlier paper\cite{BCKT}
for details about the features of the SUSY signal and SM backgrounds.

\subsection{Trilepton Channel}

To facilitate efficient computation of the signal levels in the clean
$3\ell$ channel we generated all chargino, neutralino and slepton production
subprocesses using ISAJET even though the signal is dominated by
$q\bar{q}\to \tw_1\tz_2$ production. Gluino and squark production
very seldom leads to events without any jet activity.
Following
our earlier analysis\cite{BCKT} we implement the following cuts:
\begin{itemize}
\item We require 3 {\it isolated} leptons 
within $|\eta_{\ell} |<2.5$
in each event, with $p_T(\ell_1)>20$
GeV, $p_T(\ell_2)>15$ GeV, and $p_T(\ell_3)>10$ GeV;
\item We require $\eslt >25$ GeV;
\item We require that the
invariant mass of any opposite-sign, same flavor dilepton pair not reconstruct
the $Z$ mass, {\it i.e.} we require that
$|m(\ell\bar{\ell})-M_Z|\geq 10$~GeV;
\item Finally require the events to be {\it clean}, {\it i.e.} we veto
events with jets.
\end{itemize}
SM backgrounds to the clean $3\ell$ sample mainly come from $t\bar{t}$
and $WZ$ production. Lepton isolation plays a crucial role in reducing
the top quark background. After the cuts, we are left with\cite{BCKT} 
a SM physics
background
of just 0.2~fb (mainly from $WZ$ events where $Z\to \tau\bar{\tau}$
with subsequent leptonic decays of the $\tau$s). Assuming that
detector-dependent backgrounds from particle misidentification or 
jets or photons 
faking a lepton are under control, the observation of just a handful
of such events in the current run (or at the MI) could signal
new physics. 

In Fig.~11, we show the regions of the $m_0-\mhf$ plane where the SUSY
signal is expected to be observable at the $5\sigma$ level (with a
minimum of 5 signal events) at the MI (gray squares) and at
TeV33 (white squares) for the same cases ({\it a})--({\it d}) as
in Fig.~1. We did not find any region (compatible with current experimental
constraints) where this
signal might be observable for an integrated luminosity of 100~pb$^{-1}$ only 
because our cuts in this channel were optimized for the higher Tevatron
luminosity options.
The following features are worthy of note.
\begin{enumerate}
\item We see that for small values of $m_0$ (for which sleptons are
light enough so that two body slepton-lepton decays of charginos
and neutralinos are kinematically allowed, while $\tq q$ modes
are forbidden) the MI reach
extends to $\mhf = 200-250$~GeV depending on $\tan\beta$ and the
sign of $\mu$; the gap around $m_0 \simeq 50$~GeV is where the
decay $\tz_2\to \tell_L\ell$ becomes kinematically forbidden and
where $\tz_2$ dominantly decays via $\tz_2\to \nu\tnu$. For still
larger values of $m_0$ sneutrinos become too heavy and $\tz_2 \to
\tell_R\ell$ dominates, so that the trilepton signal is again
observable. The kinematic boundaries for these various
two body slepton decay channels of $\tz_2$ are shown by the three slanting
contours in the small $m_0$ region of the Figure.
Ultimately, of course, the sleptons become
virtual, so that the neutralino decays via three-body modes. Then 
the hadronic decay of $\tz_2$ are no longer negligible, leading
to a reduction in the trilepton cross section. In fact for $m_0>150-200$~GeV,
the $3\ell$ signal falls below the observable level at the MI except
in the small $\tan\beta, \mu<0$ case ({\it a}), where the range of
$\mhf$ that can be probed slowly decreases with increasing $m_0$.

\item TeV33 should be able to probe the SUSY signal in this channel for
substantial regions of parameter space not accessible at the MI. This
should not be surprising, since the signal is essentially rate limited,
assuming that non-physics instrumental backgrounds will be in control
even in the high luminosity environment. Experiments at 
TeV33 should not only be
able to fill in most of the ``gaps'' at small $m_0$ where the signal is not
observable at the MI, but should also be able to substantially extend the
$m_0$ range where the signal might be observable. 

\item At TeV33 the clean trilepton signal may be
observable well beyond where the
``spoiler'' decay modes of $\tz_2$ become accessible if two body slepton
decays of $\tz_2$ are kinematically accessible;
the kinematic boundaries for these decays are shown by the approximately
horizontal contours.
For a limited
range of parameters, the TeV33 reach extends out to $\mhf=280$~GeV which
corresponds to $m_{\tg}=700$~GeV! There are, however, even larger ranges
of parameters where {\it this signal will not be observable at TeV33 even
if the chargino mass is at its current experimental bound.} This has been
traced to a negative interference term which leads to a large suppression
of the leptonic decay of $\tz_2$\cite{BCKT,KANE,BARTL}. It is for the
same reason that the signal is not observable all the way up to the limit
of the ``spoiler'' mode in cases ({\it b})-({\it d}). The non-observation
of a trilepton signal at TeV33 cannot, therefore, be translated into a
lower limit on the $\tw_1$ and $\tz_2$ masses.
\end{enumerate}

\subsection{Dilepton Channels}

We have also investigated the possibility of discovering supersymmetry
in the OS clean acollinear dilepton channel. Although the main
contributions to the signal might be expected to come from chargino and slepton
pair production, 
we generate {\it all} possible SUSY production
reactions and implement the following set of cuts, designed to extract signal
from these backgrounds\cite{BCKT}.
\begin{itemize}
\item We require exactly two {\it isolated} OS (either $e$ or $\mu$ ) leptons
in each event, with $p_T(\ell_1)>10$ GeV and $p_T(\ell_2)>7$ GeV, and
$|\eta (\ell ) |<2.5$. 
In addition, we require {\it no} jets, which
effectively reduces most of the $t\bar t$ background. 
\item We require $\eslt >25$ GeV to remove
backgrounds from Drell-Yan dilepton production, and also
the bulk of the background from $\gamma^*, Z\to\tau\bar{\tau}$ decay. 
\item We require $\phi (\ell\bar{\ell})<150^0$, to further reduce
$\gamma^*,Z\to\tau\bar{\tau}$ background.
\item We require the $Z$ mass cut:
invariant mass of any opposite-sign, same flavor dilepton pair not reconstruct
the $Z$ mass, {\it i.e.} $m(\ell\bar{\ell})\ne M_Z\pm 10$ GeV. 
\end{itemize}

The SM background, after these cuts, is\cite{BCKT} about 44~fb, bulk
of which comes from $WW$ production. The regions of parameter space
where the signal might be observable in this channel is shown
in Fig.~12, again for the same cases ({\it a})--({\it d}).
The horizontal (inclined) contours denote the kinematic boundaries for the
decay(s) $\tw_1 \to W\tz_1$ ($\tw_1 \to \tell_L\nu,\tnu\ell$).  
A comparison
with Fig.~11 shows that this region is a subset of the regions that
might be explored via the clean $3\ell$ channel, both at the MI
as well as at TeV33. In particular, the clean $2\ell$ channel does
not lead to an observable signal in the regions of parameter space
where the $3\ell$ signal is unobservable due to the strong suppression
of the branching fraction for leptonic $\tz_2$ decays\cite{FN4}.
Also, for the most part, the region of parameters that might be probed
at TeV33 via the clean dilepton channel may be probed at the MI via
the clean $3\ell$ channel. 

It should be possible to
confirm a signal from
$\tw_1\tz_2$ production 
by searching for OS like-flavor dileptons plus jets plus $\eslt$ 
events which arise
when the
chargino decays hadronically and the neutralino leptonically.
The invariant mass of the lepton pair in these events should
match up with the mass of the OS like-flavor dilepton pair
in the $\ell^+\ell^-\ell'$ sample of clean trilepton events.
These events generally tend to be softer than events from
the production of gluinos and squarks studied in Sec.~III.
We have examined where in SUGRA parameter space
this jetty dilepton signal might be observable, with the same
cuts as in Ref.\cite{BCKT}. Because these events are
detectable only over a subset of parameters where the clean $3\ell$
signal is observable, this channel does not lead to an improved
reach.
We will content ourselves with a
qualitative discussion of this channel.

The SM background which, after cuts, mainly comes from vector boson
pair production and Drell-Yan $\tau\bar{\tau}$ production is
about 25~fb.
The OS lepton plus jets plus $\eslt$ signal region is observable
over a subset of the regions where the trilepton signal is observable.
This region has roughly similar shape as the region in Fig.~11, but
covers only just over half as many points. For small
values of $m_0$ where the charginos predominantly decay via the two-body 
lepton-slepton
mode, these events mainly come from slepton and chargino pair
production with jets coming from QCD radiation.
For $m_0\geq
200$~GeV, the signal is observable for a rather limited range
of parameters except in case ({\it a}) where $\mhf$ values up to
100 (140)~GeV may be probed at the MI (TeV33).

It is worth mentioning that a measurement
of $\sigma (3\ell) / \sigma (\ell^+\ell^- +\jets+\eslt)$
could potentially yield a measure of the hadronic
branching fraction of $\tw_1$, provided that these events could be
separated from other SUSY sources (gluinos and squarks) which
lead to the same event topology. For very small $m_0$, we have already
mentioned that $\tw_1\tw_1$, with jets coming from QCD radiation, is the
biggest source of these events. We have checked that for larger values
of $m_0$ the $n_j \geq 2$ plus OS dilepton events, with the cuts
of Ref. \cite{BCKT} indeed come mainly from $\tw_1\tz_2$ production:
for example, for ($m_0,\mhf$)=(300~GeV, 120~GeV) in case ({\it a})
almost 90\% of these events originate in the chain $\tw_1 \to q\bar{q}\tz_1, 
\tz_2 \to \ell^+\ell^-\tz_1$. Even for the (100~GeV, 80~GeV) case, this
fraction exceeds 50\%; several other signals should be expected in such
a scenario. Although we have not attempted to explore
this, it may be interesting to examine whether the determination
of the chargino decay pattern at a high luminosity upgrade\cite{FN5} such as 
TeV33 is indeed viable. (At the LHC, one may be swamped by squark and
gluino production processes.)

\section{Summary and Conclusions}

Motivated by the recent proposal\cite{TEV2K} for upgrading the luminosity
of the Fermilab Tevatron by another order of magnitude {\it beyond the
Main Injector era}, we have examined the impact that experiments at
such a facility (should its construction prove technically and fiscally
feasible) would have on the search for supersymmetric particles. We have
used ISAJET to compute the reach of TeV33, under the optimistic
assumption of an integrated luminosity of 25~fb$^{-1}$ and compared
it with the corresponding reach of the MI, for which we assume a data sample
of 2~fb$^{-1}$. 
In our analysis, we simply assume identical
detector performance at the two facilities. We make no representation
as to whether upgrades of the detector and the electronics that would be
necessary for funtioning
in this high luminosity environment can be achieved in a timely fashion.
Thus, while our estimates for the reach of the MI are probably
realistic, 
our conclusions regarding the reach of TeV33 might be on
the optimistic side. 
We have also discussed what might be
possible from an
analysis of the current Tevatron data sample of $\sim 100$~pb$^{-1}$. 
This enables us to compare how experiments at the MI will improve
on what we can learn from the data sample of Run I of the Tevatron, and 
put in perspective the capabilities
of any upgrades that might be possible in the future.

For definiteness, we have adopted the SUGRA framework (with its
assumptions about universal scalar and gaugino masses at an ultra-high
scale) for our analysis. We have examined the multijets plus $\eslt$ plus
$n_{\ell}=1$--3 lepton
channels (Sec.~III) as well as the hadronically quiet dilepton
and trilepton channels (Sec.~IV) and
delineated regions of the $m_0$--$\mhf$ plane where the various signals
might be observable above SM backgrounds for the three values of
integrated luminosity mentioned above. Our main result is summarized
in Fig.~13, where we show the regions of SUGRA parameter space
where at least one of the SUSY signals is observable with the criteria
defined above. We show our results for the same cases ({\it a})--({\it d})
in Fig.~1. The black squares are the points that can be probed at
the ``$5\sigma$'' level with an integrated luminosity of 100~pb$^{-1}$,
while the gray (white) squares are where the signal should
be observable with a luminosity of 2~fb$^{-1}$ (25~fb$^{-1}$).
Several features are worth emphasizing:
\begin{itemize}
\item The analysis of the data from the current run will allow experiments
at the Tevatron to probe only a little beyond the current experimental
bounds. The most promising channel is the multijets plus $\eslt$ channel
from the production of gluinos and squarks; beyond 
$\mhf =100$~GeV ($m_{\tg}\simeq 300$~GeV), 
their production cross section is kinematically suppressed,
while the signals from electroweak chargino/neutralino production are still
rate limited. We stress, however, that Tevatron experiments are direct
probes of gluinos and squarks, and it is important to look
for their signals even below the hatched
region, since it is entirely possible that the assumption of gaugino
mass unification (which is crucial to translate the LEP chargino
mass bound to a bound in this plane) may prove to be incorrect.

\item Experiments at the MI should probe a significant portion of the
$m_0$--$\mhf$ plane, as can be seen from the distribution of gray
squares in Fig.~13. Here, the important contributing channels are the multijets
plus $\eslt$ channel and, especially in the small $m_0$ region, the
clean $3\ell$ channel. We see that the region of the $m_0-\mhf$ plane
that can be explored is sensitive to both $\tan\beta$ and $\sgn\mu$: for
favourable values of these parameters, the experiments may probe
$\mhf$ as large as 250~GeV if $m_0 \leq 150$~GeV via the clean
trilepton channel; but for somewhat larger values of $m_0$, we see 
from Figs.~13~({\it b})--({\it d}) that there
may be no SUSY signal at the MI even if charginos are at their current
experimental bound.

\item Assuming an integrated luminosity of 25~fb$^{-1}$, we see that
experiments at TeV33 may be able to substantially expand the region
where a SUSY signal might be observable. Generally speaking, the
reach is most enhanced in the rate-limited $3\ell$ channel (where
the gain is proportional to the integrated luminosity ($L$), unlike
the background-limited multi-jet channels where the statistical significance
of the signal improves only as $\sqrt{L}$). Nevertheless, it is
important to note that there are significant ranges of parameters
which cannot be efficiently probed at TeV33 even if charginos are
just beyond the reach of LEP2 (recall that LEP2 will probe the
region $m_{\tw_1} \leq 80-90$~GeV well before the TeV33 commences 
operation). It will thus be difficult to obtain an unambiguous lower
bound on sparticle masses if no SUSY signal is observed in experiments 
at TeV33.

\item We also  see from Fig.~13 that experiments at the MI and TeV33
will substantially expand the SUSY reach of the Tevatron. 
An important virtue of TeV33, however, is that there is an observable
SUSY signal in several channels over a substantial portion of parameter
space where SUSY should be detectable. 
Furthermore, if a signal for supersymmetry is found at the MI, then the
order of magnitude increase in collider luminosity at TeV33 
should allow more detailed information about the underlying SUSY parameters
to be extracted from the event sample.

\end{itemize}

Our analysis was performed within the minimal framework
where the conservation of $R$-parity is assumed. If instead
$R$-parity is violated by baryon number non-conserving operators so
that the LSP decays hadronically, the $\eslt$ as well as the 
multilepton signals are greatly diminished, and the reach may
be significantly smaller\cite{REDUCE} than that outlined in Fig.~13.

At this point it is worth recalling that in order for
supersymmetry to ameliorate the fine-tuning problem of
the SM, sparticles cannot be arbitrarily heavy\cite{REV}:
qualitatively, one requires that sparticles are not much
heavier than the weak scale. Several authors\cite{GIUDICE,DIEGO}
have attempted to quantify this and obtained {\it upper}
limits on sparticle masses. While these bounds are admittedly
subjective, they could be regarded as providing rough benchmarks
for future facilities; {\it e.g.} Anderson and Casta\~no\cite{DIEGO}
have argued that the most favoured region from this point
of view is where $\mhf \alt 150$--200~GeV, $m_0 \alt 200$--300~GeV.
It is also interesting to note that the lightest neutralino would
be an acceptable mixed dark matter candidate if SUGRA parameters happen
to be in this range\cite{BRHLIK}. These arguments suggest that
experiments at the MI and TeV33 would probe some of the most promising
regions of SUGRA parameter space.
We believe, however, that while fine-tuning and cosmological considerations
are indeed suggestive, the upper bounds on sparticle masses that are
obtained from these should be regarded as qualitative, and that a sufficient
``safety margin'' should be allowed for any experiment that is designed
to decisively confirm or exclude weak scale supersymmetry.
This
is only possible at hadron supercolliders such as the LHC
where it is possible to probe the $m_0$--$\mhf$ plane over the region
$m_0 \leq 1.5$~TeV, $\mhf \leq 800$~GeV\cite{LHC},
or at future electron-positron colliders operating at $\sqrt{s}\geq
1$--1.5~TeV,
where it should be possible to probe chargino and slepton masses
up to the beam energy. 


%
\acknowledgments
This research was supported in part by the U.~S. Department of Energy
under contract number DE-FG05-87ER40319, DE-FG03-91ER40674,
DE-AC02-76CH00016, and DE-FG-03-94ER40833. 
%
%
%
%

\begin{figure}
\caption[]{Contours of squark (dashed), gluino (solid) and chargino
(dotted) masses in the 
$m_0$-$\mhf$ plane of the minimal SUGRA model. Frames are shown for
{\it a}) $\tan\beta =2,\ \mu <0$, {\it b}) $\tan\beta =2,\ \mu >0$,
{\it c}) $\tan\beta =10,\ \mu <0$, and {\it d}) $\tan\beta =10,\ \mu >0$.
We take $m_t=170$ GeV and $A_0=0$. The bricked regions are excluded by 
theoretical constraints discussed in Ref.\cite{LHC} 
while the shaded regions are excluded by experiment.}
\end{figure}
\begin{figure}
\caption[]{Component SM background cross sections as a function of the
cut parameter $\ETC$ defined in the text for multijet plus a)~$\eslt$,
b)~$1\ell+\eslt$, c)~OS dileptons$+\eslt$, d)~SS dileptons$+\eslt$,
e)~$3\ell+\eslt$ and f)~tagged $B+\eslt$ event topologies with cuts
as defined in Sec.~III of the text. The backgrounds that we have computed
are $t\bar{t}$ (long-dashed), QCD (long-dashed-single-dotted),
$W+jets$ (long-dashed-double-dotted),
$Z+jets$ (long-dashed-triple-dotted), $WW$ (short-dashed),
$WZ$ (short-dashed-dotted) and $ZZ$ (short-dashed-double-dotted). The
sum of these backgrounds is shown as the solid curve.}
\end{figure}
\begin{figure}
\caption[]{ Variation of the SUSY signals and total SM background with
$\ETC$ for the same event toplogies as in Fig.~2. We have fixed
$\mhf=120$~GeV, $\tan\beta=2$,
and chosen $A_0=0$ and $\mu<0$ with $m_0=100$~GeV (solid),
$m_0=200$~GeV (dashed) and $m_0=800$~GeV (dotted). In the other two cases,
we take $\mu>0$ and fix $A_0=-400$~GeV (dot-dashed) and $A_0=-430$~GeV 
(dot-dot-dashed). The SM background level is denoted by crosses.}
\end{figure}
\begin{figure}
\caption[]{Regions of the $m_0-\mhf$ plane where the multijets plus
$\eslt$ signal
is observable at a 2~TeV $p\bar{p}$ collider
according to the criteria discussed in Sec.~III of the text
for the same choices of parameters as in Fig.~1. We consider
three values for the integrated luminosity:
100~pb$^{-1}$
corresponding to Run I of the Tevatron (black squares),
2~fb$^{-1}$ which is expected to be accumulated at
the Main Injector (gray squares) and  25~fb$^{-1}$ to be accumulated
at the proposed TeV33 (white squares). The bricked and hatched regions
are the same as in Fig.~1.}
\end{figure}
\begin{figure}
\caption[]{The same as Fig.~4 except for the multijet plus $1\ell+
\eslt$ channel.}
\end{figure}
\begin{figure}
\caption[]{The same as Fig.~4 except for the multijet plus OS dilepton
$+\eslt$ channel.}
\end{figure}
\begin{figure}
\caption[]{The same as Fig.~4 except for the multijet plus SS dilepton
$+\eslt$ channel.}
\end{figure}
\begin{figure}
\caption[]{The same as Fig.~4 except for the multijet plus
$3\ell+\eslt$ channel.}
\end{figure}
\begin{figure}
\caption[]{The same as Fig.~4 except for the multijet plus
tagged $B+\eslt$ channel.}
\end{figure}
\begin{figure}
\caption[]{The $A_0$ dependence of the SUSY signal
cross sections in the six channels in Fig.~3. for $m_0=100$~GeV,
$\mhf=120$~GeV, $\ETC=15$~GeV for $\tan\beta =2$ and 10 and
for either sign of $\mu$.}
\end{figure}
\begin{figure}
\caption[]{The same as Fig.~4 except for the clean, {\it i.e.} jet-free
$3\ell+\eslt$ channel discussed in Sec. IV. The three slanting contours
(from left to right) mark the boundaries of the regions where the
two body decays $\tz_2 \to \tell_L\ell, \tnu\nu$ and $\tell_R\ell$
become forbidden, whereas the (almost) horizontal contours mark the
corresponding boundaries for the ``spoiler'' decays $\tz_2\to Z\tz_1$ or
$H_{\ell}\tz_1$. Outside the 
hatched region, we found no points where
this signal would be observable from an analysis of the data from Run I of
the Tevatron. }
\end{figure}
\begin{figure}
\caption[]{The same as Fig.~11 except for the clean dilepton channel. The
slanting contours, from left to right,  mark the kinematic boundary for the
decays $\tw_1 \to \tell_L\nu,\tnu\ell$, while the roughly horizontal line
marks the boundary for the decay $\tw_1\to W\tz_1$.}
\end{figure}
\begin{figure}
\caption[]{The cumulative reach for supersymmetry
of the Tevatron and its upgrades
via any of the channels in Figs.~4-12 with the same labelling as
in Fig.~4.}
\end{figure}


\begin{references}
\bibitem{FN1} For instance, while Ref.\cite{LOPEZ} uses a real
detector (CDF) simulation for the analysis of the background
to the clean trilepton signal, the background to the $\etmiss$
channel is evaluated by computing the background from $Z\to\nu\nu+jets$
and multiplying it by five. In contrast, Ref.\cite{BCKT} has focussed
on only the di and trilepton signal using a toy simulation.
Unlike these
analyses which use ISAJET, the work of Ref.\cite{KANE} is based on
PYTHIA. These authors have studied the reach in the clean trilepton
as well as in the $\etmiss$ channel but we will see that our estimates
of SM backgrounds are frequently considerably larger than those in
Ref.\cite{KANE}. Finally, we note that although all three groups have
performed their studies within the SUGRA framework,
it is difficult
to directly compare the reach that they obtain
because of the way that the
three groups choose to present the results. In Ref.\cite{LOPEZ,KANE}, random
points are generated across the entire allowed parameter space, but then
the experimental cross sections are shown as a function of some particular
sparticle mass (and all information about the underlying SUGRA parameters
is lost), while in Ref.\cite{BCKT} the results are shown in the $m_0-\mhf$
plane, for particular choices of $A_0$, $\tan\beta$ and $sgn(\mu )$.
\bibitem{FN2} The current version of ISAJET does not include some of the
effects from Yukawa couplings of $b$-quarks and $\tau$ leptons. Because
these may become important for large values of $\tan\beta$, we have
not included cases with $\tan\beta \simeq \frac{m_t}{m_b}$ in our
study.
\bibitem{FN4} This conclusion differs from that in our previous
study\cite{BCKT} where we found (see Fig.~8 of that paper) that
the clean $2\ell$ signal might be observable at TeV33 within
the ``hole region'' for the trilepton signal. The range of parameters
where this occured is now excluded by the LEP1.5 constraint on the chargino
mass.
\bibitem{FN5} After cuts, $\sigma(3\ell)$ ($\sigma(\ell^+\ell^- +$jets))
is 5~fb and 30~fb (17~fb and 88~fb) for the two cases just mentioned.
A measurement of the branching fraction thus does not appear to be feasible 
with the MI data sample.
\bibitem{REDUCE} We are not aware of any detailed calculations of
the SUSY reach of TeV33 in these scenarios. At the MI, it is possible
that there may be no observable signal even if gluinos are relatively
light and squarks are heavy as reviewed in Ref.\cite{RPV}. The statistical
significance of the signal should be better at TeV33, but not dramatically
so since the channels are mostly background-limited. Of course, in the
other case where the LSP decays only into $e$ or $\mu$ via $R$-parity
interactions that violate lepton number, it should be possible to
probe gluino masses of 500-600~GeV even at the Main Injector\cite{RPV}.
\end{references}
\end{document}